\documentclass[12pt]{article}

\usepackage[utf8]{inputenc}
\usepackage{amsmath, amsthm, amssymb, color, mathbbol}
\usepackage{graphicx}
\usepackage{makecell}
\usepackage{multicol}

\usepackage[table,xcdraw]{xcolor}
\usepackage{tikz}
\usepackage{array,amscd}
\usepackage{amsfonts}
\usepackage[T1]{fontenc} 
\usepackage{enumerate}
\usepackage{appendix}
\usepackage[arrow, matrix, curve]{xy}
\usepackage{booktabs}
\usepackage{siunitx}
\usepackage{multirow}
\usepackage[font=small,labelfont=bf]{caption}

\usepackage{tikz}
\usetikzlibrary{snakes}
\usetikzlibrary {shapes}
\usetikzlibrary {arrows}
\usetikzlibrary {positioning}
\usetikzlibrary {calc}

\oddsidemargin \evensidemargin
\usepackage{enumitem}
\setenumerate{leftmargin=*}
\allowdisplaybreaks[3]

\definecolor{darkgreen}{rgb}{0,0.4,0}
\definecolor{MyBlue}{rgb}{0,0.08,0.7} 
\definecolor{MyRed}{rgb}{0.85,0.08,0}

\makeatletter
\renewcommand*\env@matrix[1][\arraystretch]{%
  \edef\arraystretch{#1}%
  \hskip -\arraycolsep
  \let\@ifnextchar\new@ifnextchar
  \array{*\c@MaxMatrixCols c}}
\makeatother

\usepackage{authblk}
\usepackage{subfig}
\usepackage{threeparttable}
\usepackage{footnote}
\usepackage{amssymb}
\usepackage{booktabs}

\usepackage[utf8]{inputenc}
\usepackage[english]{babel}
 \usepackage{xcolor}
   
\usepackage{amsthm}
\usepackage{afterpage}

\newcommand{\indep}{\perp \!\!\! \perp}

\usepackage[linesnumbered,ruled,vlined]{algorithm2e}

\providecommand{\keywords}[1]{\textbf{\textit{Keywords---}} #1}

\newcommand\blankpage{%
    \null
    \thispagestyle{empty}%
    \addtocounter{page}{-1}%
    \newpage}

\usepackage[utf8]{inputenc} 
\usepackage[T1]{fontenc}    
\usepackage{hyperref}       
\usepackage{url}            
\usepackage{booktabs}       
\usepackage{amsfonts}       
\usepackage{nicefrac}       
\usepackage{microtype}      
\usepackage{lipsum}
\usepackage{graphicx}
\usepackage[numbers]{natbib}

\title{\textbf{Modeling "Equitable and Sustainable Well-being" (BES) using Bayesian Networks: \\A Case Study of the Italian regions}}

 \author{Federica Onori, Giovanna Jona Lasinio}

\date{}

\begin{document}

\maketitle

\theoremstyle{definition}

\theoremstyle{remark}

 \affil[1]{Department of Statistical Sciences,
   University of Rome La Sapienza,
      Rome}
 \affil[*]{Corresponding author:  Federica Onori ,   onori.federica@gmail.com}



\maketitle

\begin{abstract}
Measurement of well-being has been a highly debated topic since the end of the last century. While some specific aspects are  still  open issues, a multidimensional approach as well as the  construction of shared and well-rooted systems of indicators are now accepted as the main route to  measure  this complex phenomenon. A meaningful effort, in this direction, is that of the Italian "Equitable and Sustainable Well-being" (BES) system of indicators, developed by the Italian National Institute of Statistics (ISTAT) and the National Council for Economics and Labour (CNEL). The BES framework comprises a number of atomic indicators measured yearly at regional level and reflecting the different domains of well-being (e.g. Health, Education, Work \& Life Balance, Environment,...). 
In this work we aim at dealing with the multidimensionality of the BES system of indicators and try to answer three main research questions: I) What is the structure of the relationships among the BES atomic indicators; II) What is the structure of the relationships among the BES domains;
III) To what extent the structure of the relationships reflects the current BES theoretical framework.
We address these questions by implementing Bayesian Networks (BNs), a widely accepted class of multivariate statistical models, particularly suitable for handling reasoning with uncertainty. Implementation of a BN results in a set of nodes and a set of conditional independence statements that provide an effective tool to explore associations in a system of variables. In this work, we also suggest two strategies for encoding prior knowledge in the BN estimating algorithm so that the BES theoretical framework can be represented into the network.
\end{abstract}

\blankpage{}

\section{Introduction}

The measurement and definition of well-being, as referred to both individuals and societies, is a long-time debated topic  that in  recent times increasingly captured the attention of stakeholders and policy-makers \citep{sen1999commodities}. This interest has been mostly motivated by the broadening of well-being definition, going further than the usual economic status \citep{wesselink2007measurement}. Hence the need of finding measures other than the economic ones that could reflect the level of societal progress. In a society funded on the awareness that "what is being measured" has direct consequences on "what is done" \citep{commission2009report}, it is of primary importance to develop proper  measurement tools that are capable to catch all relevant features of a complex phenomenon and such that assumptions and ideas are easily and clearly included. Such tools will allow  policy-makers and stakeholders to make correct, transparent and efficient decisions.

Since $2001$, the Organization for the Economic and Co-operation Development (OECD) recognized the importance of finding a proper way to measure social progress  and promoted the international debate around its definition and measurement. With the "Istanbul Declaration", jointly signed in $2007$ by the European Commission, OECD, Islamic Conference Organization, United Nations Development Program (Undp) and World Bank,  a first international consensus was reached on the necessity of measuring social progress in every Country embracing a  vision capable of going beyond the use of the Gross Domestic Product (GDP). The most relevant contribution, in these terms, is that of the "Final report of Stiglitz-Sen-Fitoussi  Commission" \citep{commission2009report}, where several recommendations are made. Firstly, the authors suggest to draft the focus from production to available income and consumption and to pay special attention to the  distribution of wealth in the population. Secondly, well-being should be conceived as a \emph{multidimensional} phenomenon, and its measurement should contemplate citizen's subjective features, and include indicators specifically meant to capture economic, social and environmental sustainability. Other tangible initiatives soon followed, all aiming at establishing  work directions towards the implementation of effective measurement of  well-being. Among these, a round table arranged on the topic during G20 in Pittsburgh (2009), with the specific intent of extending the measurement of economic development in order to include social and environmental dimensions; and the European Commission's initiative "Not only the GDP", through which the European Commission and the member states committed to elaborate on specific aspects: 1) to complement GDP with social and environmental indicators; 2) to produce timely and effective measurements of the social and environmental aspects; 4) to give more precise measurements of wealth distributions and inequalities; 3) to come up with a possibly harmonized European table to evaluate social development. Noteworthy, this is also the framework in which looking at the Sustainable Development Goals (SDGs) defined in the 2030 Agenda signed by the United Nations in 2015, and aiming at defining goals and targets for all countries to reach, specifically highlighting the importance of sustainability as an inescapable characterization of the process of development. 

Multiple and diverse  initiatives trying to implement the above-mentioned tasks and challenges followed, at both national and international levels. The "Canadian Index of Well-being" (Ciw), the "Measures of Australia’s Progress", the Buthan's "Gross National Happiness Index", the "Measuring National Well-being" project in the United Kingdom and the "Final Report adopted by the European Statistical System Committee", from the "Sponsorship Group on Measuring Progress, Well-being and Sustainable Development", are just some of them. Among the most ambitious initiatives attempting at measuring well-being, the Italian "Equitable and Sustainable Well-being" (BES) project finds its own place. The BES project started in  $2010$, when a  steering committee was formed by the joint effort of the Italian National Institute of Statistics (ISTAT) together with the National Council for Economics and Labour (CNEL) with the objective of elaborating on a system of indicators that would achieve a proper measurement of the Italian well-being. This experience was soon characterized by being widely shared among all interested social parties: field's experts, stakeholders, policy officials, statisticians, civil society's delegates and the citizens (through citizen's consultations). This choice was motivated by the consideration that the concept of well-being unavoidably changes across time, places and cultures, and thus it can only be defined together with all relevant social actors, fact that gives BES a remarkably democratic legitimacy. The BES system of indicators represents a unique chance to monitor all relevant aspects of well-being in the Italian context and this work aims at investigating possible relationships among the BES atomic indicators and the BES domains. In order to embrace the inherent multidimensionality of the BES system of indicators, the multivariate class of models of Bayesian Networks (BNs) has been implemented and different geographic scales have been tested in order to find the best grouping of the Italian regions that would allow for these relationships to emerge. The basic idea of BNs is to combine probability and graph theory by encoding the set of \emph{conditional independence relationships} among a set of variables in a graphical structure that can be easily read and interpreted. The graphical structure is made up of nodes and edges between pairs of node. Each variable is represented by a node in the graph and a missing edge between two nodes corresponds to a conditional independence relationship between the two variables. Compared to other classes of models, BNs seem to present some advantages. Unlike, for example, classical multiple regression models, BNs can also return indirect relationships mediated by third  variables and do not need to calibrate the model by specifying one or more target variable, standing out as a more suitable tool to study a phenomenon from a \emph{systemic} point of view. If compared to other multivariate statistical models like, for example, factor analysis, BNs have the advantage of being able to build an entire system of relationships to be possibly used as a decision making tool. Moreover, if compared with Neural Networks (NNs), BNs can explicitly formalize the set of relationships among variables and, as a consequence, results are generally highly interpretable. 
In literature, the BES dataset has been inspected from several points of view: cluster analysis \citep{monte2017benessere}, factor analysis \citep{chelli2015comparing,chelli2016assessing, mazziotta2019use}, regression models \citep{istat2018benessere}, partial least squares path modeling \citep{davino2017quantile}. To the best of the authors' knowledge, this is the first time that the class of models of Bayesian Networks is considered for the BES dataset, even though its usage in the field of well-being was already suggested \citep{maggino2012measuring, ceriani2016multidimensional}. In this work, we implement Bayesian Networks for the BES atomic indicators in order to answer three main research questions:
   \begin{enumerate}
     \item What is the structure of the relationships, in terms of conditional independence statements, among the \emph{BES atomic indicators}? 
     \item What is the role that the different \emph{BES domains}, as they are currently defined and formalized in the BES framework, play in the whole BES system of indicators and how do they relate to each other?
     \item To what extent the structure of the relationships reflect the current BES theoretical framework? 
     \end{enumerate}
     The first two questions mainly refer to the possibility of identifying mechanisms that can guide resources' allocation in policy decision making. This specifically instantiates in pinpointing either BES atomic indicators or domains having a central role in spreading information over the network as well groups of either indicators or domains that strictly relate to each other. In order to distinguish the level of the analysis, by looking at either BES atomic indicators or BES domains of well-being, we do exploit the flexibility of Bayesian Networks in formalizing prior information and suggest two different prior strategies, with Strategy $1$ focusing exclusively on relationships among the BES atomic indicators and Strategy $2$ devoted to capture interrelations among the BES domains.
     In order to find the most suitable geographic scale through which looking at the above mentioned relationships, four different grouping options for the Italian regions were tested: regional (NUTS 2 level), macro-areas (NUTS 1 level: NW-NE-C-S-I), North/South and national. Considerations derived from implementing both prior settings will also allow to answer the third research question, which basically refers to the opportunity of highlighting some future directions for the BES system of indicators in light of the relationships arising from the network.

The work will proceed as follows. In section $2$  we give an in-depth overview of the "Equitable and Sustainable Well-being" (BES) framework and report on the most important characteristics of the data. In section $3$ we briefly review the statistical tools of the Bayesian Network's methodology. In section $4$ we present results from implementation of the Bayesian Networks on the BES dataset. We conclude, in Section $5$, by commenting on the obtained results and suggesting some directions for future works.

\section{Well-being as a multidimensional phenomenon: a change in perspective in the Italian context} \label{second}

In order to reach a definition of well-being that could be appropriate for the Italian context, the starting point was the theoretical framework elaborated by OECD \citep{hall2010framework}, according to which specific domains and dimensions for the phenomenon are identified. In agreement with this framework, a preliminary distinction is made between 
Human well-being and Ecosystem well-being. Within the former, a further separation is also made between Individual and Societal well-being, both relying on three supporting pillars: culture, economy and governance. These pillars are not meant as final goals but rather as means to reach both Individual and Societal well-being. On the other hand, the Ecosystem well-being accounts for one domain, namely the one of environmental conditions and, evidently, important exchanges intervene between Human well-being and Ecosystem well-being, in terms of ecosystem's services and resources management. An important perspective through which considering this theoretical framework for the societal progress is that of its distribution, both across regions (space distribution) and generations (time distribution). The regional distribution is what has been identified as the \emph{equity} aspect that the definition of well-being necessarily must consider, hence the need  for well-being to be \emph{equitable}. While the time (or inter-generational) distribution refers to the fact that the current well-being should guarantee the possibility to be maintained and augment also for future generations  \citep{oposa1994defense, gardiner2011perfect}, from which the need  for well-being to be also \emph{sustainable}.

Grounded on this theoretical bases, 
the \emph{BES framework} consists of two different sets of domains. The first one comprises the \emph{outcome} domains, i.e. those domains that directly affect human and environmental well-being. A total of $12$ outcome domains have been pinpointed, namely: Health, Education,  Work \& Life Balance (Quality),  Work \& Life Balance (Employment), Economic Well-being (Min. Conditions), Economic Well-being (Income), Social Relationships, Security (Predatory Crimes), Security (Murders), Subjective Well-being, Landscape \& Cultural Heritage, Environment. Conversely, the elements of the second group are referred to as \emph{contextual} (or instrumental) domains, and they incorporate the functional elements that allow well-being to express, as regards both society and the environment. Within the BES framework, $3$ domains belong to this second group: Politics \& Institutions,  Research, Innovations \& Creativity, Quality of Services. For each domain, a number of atomic indicators are selected in order to monitor all relevant features.
At present, a total of $129$ atomic indicators have been identified and, among these, $61$ indicators are aggregated on a yearly basis into domain composite indicators by means of the \emph{Adjusted Mazziotta-Pareto Index} (AMPI) \citep{mazziotta2016generalized}, following criteria of timeliness, availability of the time series and capability at representing the different features of each domain. The Mazziotta-Pareto Index (MPI) works under the hypothesis that atomic indicators to by synthesized are not exchangeable and all have the same relevance. In its \emph{adjusted} version, an ad hoc normalization procedure is implemented in order to allow comparability across years \citep{de2009composite}. Henceforth, we will refer to the $61$ indicators considered for building the composite indicators as the \emph{BES atomic indicators}. Domains and corresponding atomic indicators used for the synthesis, are reported in Table \ref{tab:bes1} and Table \ref{tab:bes2} in Appendix. For all but three domains, one single composite indicator is computed, while for  Work \& Life Balance, Economic Well-being and Security, two composite indicators are considered, to better represent heterogeneous attributes within the same domain \citep{istat2018benessere}.

The measurement model used for the BES framework follows the so called \emph{formative} approach, since indicators are deemed to \emph{define} the latent trait representing the phenomenon, contrary to the \emph{reflexive approach}, according to which the phenomenon reflects into indicators \citep{maggino2017developing,diamantopoulos2001index, diamantopoulos2006formative}. This distinction has important consequences on the measurement
model specification in terms of the relationship between indicators and the latent trait. In fact, within a formative approach indicators are not exchangeable, and a different set of indicators would result in a different operational definition (and thus a different value) for the latent trait being measured \citep{blalock2018causal}. Furthermore, indicators are not required to be correlated to each other (i.e. no internal consistency in terms of correlation is required) and, if any correlation shows up, it is not strictly explained by the model itself. For what the error term concerns, it only has a disturbing role within the measurement model. On the contrary, according to the reflective approach, indicators are exchangeable, must be correlated, the correlation is explained by the model and the error term plays a role in explaining the indicators' variance. It is worth mentioning that a long-term, yet on-going debate pertain to scholars of different views, about the superiority or even the suitability of one approach with respect to the other  (see for example \citep{howell2007reconsidering, wilcox2008questions, bollen2017defense}.

Some important characteristics of the BES atomic indicators employed to compute composite indicators immediately stand out, and we will briefly consider the main ones \citep{istat2015benessere}. First of all, the atomic indicators do not yearn for an exhaustive representation of the considered domain (for which a huge number of indicators should instead be envisioned), but rather aim at catching those aspects that mainly characterize the domain, while ensuring a timely monitoring and the proper geographic breakdown. Secondly, both objective and subjective indicators are used, in the belief that citizens' perceptions and opinions can complement knowledge about well-being, providing pieces of information that could not be added otherwise. Thirdly, all atomic indicators must encounter at least the regional disaggregation (NUTS2) level, possibly allowing the monitoring of the phenomenon in all its aspects also at a more local level. Lastly, indicators should allow for an analysis disaggregated at the level of some relevant structural variables (sex, age, education, social status, etc...)

In this work, we deal with $61$ BES atomic indicators currently used to build composite indicators for each domain and we will sometimes refer to this set of indicators as the \emph{BES dataset} (Table \ref{tab:bes1} and Table \ref{tab:bes2} in Appendix). The Italian National Institute of Statistics (ISTAT) updates the time series of the BES atomic indicators at the regional level on a yearly basis, starting from $2010$. Therefore, the available BES dataset comprises values for the $61$ atomic indicators for the $20$ Italian regions for a period of eight years, from $2010$ to $2017$, for a total of $160$ observations. Although it is usually common to consider the BES atomic indicators as synthesized into composite indicators, one for each domain (see Section \ref{second}), in this work information will be kept at the atomic indicators level, with the twofold intention of not (potentially) loosing information along the synthesis process and deal with variables whose meaning is maximally clear and transparent.
However, it will also be shown how the theoretical framework of the $61$ atomic indicators organized into $15$ well-being domains can conceivably be incorporated into the BN learning scheme and what kind of information it is possible to gain out of it.

\section{Methodology}

\subsection{Bayesian Networks} \label{BN}
Bayesian Networks models (or Bayesian Networks, shortly BNs) are a class of multivariate statistical models that allow for the representation of probabilistic relationships among a set of variables by means of a graph \citep{ lauritzen1996graphical, pearl2014probabilistic, neapolitan2004learning, koller2009probabilistic}. Bayesian Networks have been extensively applied in a variety of fields, from
genomic and gene expression analysis \citep{friedman2000using, rodin2005mining,  scutari2014multiple} to official statistics \citep{sebastiani2001use, penny2004using},
ecology \citep{milns2010revealing,hradsky2017bayesian},
fault diagnosis \citep{helldin2009explanation},
healthcare \citep{lappenschaar2013multilevel, van2014learning}  and climate data applications
\citep{ebert2012causal, molina2013dynamic, vitolo2018modeling}.

The present section aims at providing the fundamental elements for understanding Bayesian Networks. For a complete theoretical treatise, we refer to the well established literature \citep{lauritzen1996graphical, pearl2014probabilistic, koller2009probabilistic, whittaker2009graphical, scutari2014bayesian, cowell2006probabilistic}.

Given a set $\mathbf{X}=(X_{1}, \dots, X_{n})$ of random variables, a Bayesian Network is defined by specifying
\begin{itemize}
    \item A graph $G= (V, E)$, where $V$ is a set of vertices (or nodes), one for each random variable in $\mathbf{X}$ (nodes and random variables are used interchangeably), while $E \subseteq V \times V$ is a set of edges (links) connecting pairs of different nodes. For a BN to be properly specified, the graph $G$ must be a \emph{directed} and \emph{acyclic} (shortly, a DAG).
    \item A multivariate probabilistic distribution defined over $\mathbf{X}$, called the \emph{global} distribution. The univariate distributions associated to each $X_{i}$ are called  \emph{local} distributions. 
    
\end{itemize}

Let $\alpha,\beta \in V$ be two vertices, we write $\alpha \to \beta$ if an arrow exists going from $\alpha$ to $\beta$. In this case we call $\alpha$ the \emph{parent} node and $\beta$ the \emph{children} node. The set of parents of a node $\beta$ is denoted as $pa(\beta)$ and the set of children of a node $\alpha$ is denoted as $ch(\alpha)$. A \emph{path} from $\alpha \in V$ and $\beta \in V$ is a sequence $\alpha = \alpha_{0}, \dots, \alpha_{n} = \beta
$ of distinct vertices such that $(\alpha_{i-1}, \alpha_{i}) \in E$, for all $i= 1,\dots,n$. If there is a path from $\alpha$ to $\beta$, we say that $\alpha$ \emph{leads} to $\beta$ and we write $\alpha \mapsto \beta$. On the contrary, those nodes $\beta$ for which no path exists leading from $\alpha$ to $\beta$ are said to be \emph{non-descendants}. A central property for Bayesian Networks is the so called the (local) \emph{Markov property},
 according to which each variable $X_{i}$ is conditionally independent of its non-descendants given its parents. As a consequence of both the Markov property and the chain rule, the joint distribution can be conveniently factorized into a product of marginal/conditional distributions \citep{lauritzen1996graphical}, which is certainly one of the most relevant features of Bayesian Networks. In fact, it allows algorithms to scale to fit high-dimensional data and avoid incurring in the so called \emph{curse of dimensionality} \citep{ wainwright2008graphical, jordan1998learning, kolar2013uncovering}. 
Ultimately, the DAG being acyclic guarantees that factorizations are indeed well defined.

The possibility of establishing a link between graphical structures and probabilistic properties is made possible by defining a suitable concept of \emph{graphical separation} for subsets of nodes in a graph (the so called \emph{$d$-separation}, and then relating it to the well known conditional independence property for a set of random variables (for more details on this aspect, please refer to \citep{lauritzen1996graphical,koller2009probabilistic}). Remark that perhaps the most informative feature of an edges set is the absence of an edge: if two nodes $X_{i}$ and $X_{j}$ are not connected by an edge, then the two random variables are either marginally or conditionally  independent  (i.e. they are independent when conditioning on a subset of the remaining variables). On the other hand, the presence of an edge connecting two nodes in the graph reflect possible conditional dependencies between the two variables. In fact, after a Bayesian Network is estimated from a real dataset, \emph{an evaluation step} performed by field experts should follow, in order to better understand the final results \citep{ebert2015using}. Some of the connections could for example be confirmed by the scientific literature on the relative field and, if that is the case, the edges could indeed be thought of as corresponding to specific \emph{mechanisms in action}. On the contrary, if literature cannot confirm a connection, this could potentially shed light on (maybe new) hypothesis to be tested by future works.

\subsubsection{Fundamental Connections} \label{fundamental_connections}
\begin{figure} 
  \centering
\includegraphics[width=10 cm,height=8 cm,keepaspectratio]{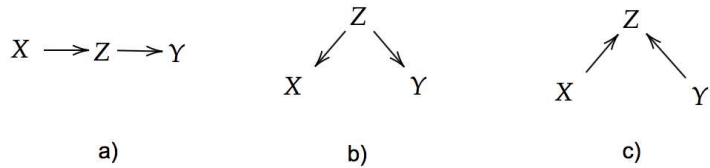}
\caption{Bayesian Networks' three fundamental connections: a) serial connection; b) diverging connection; c) converging connection.}
  \label{fig:struct}
\end{figure}

It is possible to think of the graphical structure of a BN as it was built upon some "building blocks" known as the \emph{fundamental connections} \citep{whittaker2009graphical, scutari2014bayesian}. These are the only possible configurations for three nodes and two edges and they are (see Figure \ref{fig:struct})  \begin{itemize}
\item \emph{Converging} connections: structures of the type $X \to Z \gets Y$. A converging connection where $X$ and $Y$ are not connected is said to be a \emph{$v$-structure}.
    \item \emph{Serial} connections: structures of the type $X \to Z \to Y$.
    \item \emph{Diverging} connections: structures of the type $X \gets Z \to Y$.
    
\end{itemize}

The three connections represent different probabilistic relationships by means of different graphical structures. In converging connections, the graph is associated with the marginal independence statement $X \indep_{P}Y$ while, when conditioning on the value of the variable $Z$, $X$ and $Y$ become conditionally dependent. Furthermore, the joint distribution factorizes as $P(X,Y,Z) = P(X)P(Y)P(Z|X,Y)$. In serial and diverging connections, the missing edge between nodes $X$ and $Y$ corresponds to the conditional independence statement $X \indep_{P}Y|Z$ and, for both kinds of connections, the joint distribution factorizes as $P(X,Y,Z) = P(X)P(Z|X)P(Y|Z) = P(Z)P(X|Z)P(Y|Z)$. 
Therefore serial and diverging connections - albeit different graphical structures - encode the same conditional independence relationships and thus reveal the same factorization for the joint distributions. Such graphical structures are said to be in the same \emph{equivalence class} \citep{frydenberg1990chain, verma1991equivalence}.

\subsubsection{In-degree, Out-degree and Markov Blanket size} \label{MB}
\begin{figure} 
  \centering
\includegraphics[width=7 cm,height=6 cm,keepaspectratio]{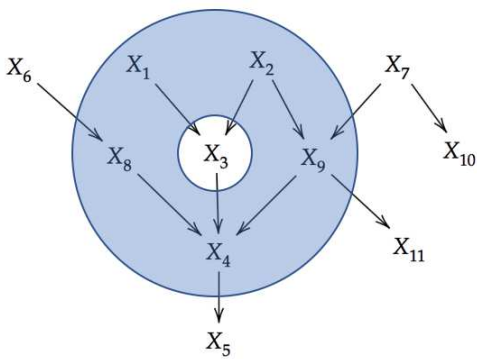}
\caption{A directed acyclic graph (DAG) with eleven nodes.
\normalsize{The Markov blanket of node $X_{3}$ comprises the nodes in the blue region: its parents $\{X_{1}, X_{2}\}$, its child $\{X_{4}\}$ and its child's other parents $\{X_{8}, X_{9}\}$. }As per the local Markov property, the global probability distribution factorizes as
\footnotesize{$P(\mathbf{X}) = P(X_{1})P(X_{2})P(X_{6})P(X_{7})P(X_{8}|X_{6})P(X_{3}|X_{1}, X_{2})P(X_{9}|X_{2}X_{7})P(X_{10}|X_{7})P(X_{4}|X_{8}X_{3}X_{9})P(X_{11}|X_{9})P(X_{5}|X_{4})$}.}
 \label{fig:mb}
\end{figure}

When reading information out of a network, it is relevant to quantify the degree of connection of each node to the whole network which in turn allows to characterize the different roles that the nodes fulfill.  There are different ways for quantifying the degree of connection of a node and the in-degree, out-degree and Markov blanket size are among the more explicative and more easily interpretable ones. The \emph{in-degree} of a node is the number of edges pointing to that node, while the \emph{out-degree} of a node is the number of edges departing from that node. In order to define the Markov blanket (and thus the Markov blanket size) of a node, we should refer to the concepts introduced in the last paragraphs. The joint distribution and density factorization derived from the Markov property, and discussed in the previous paragraph, provide a very convenient way to describe (and define) a BN. Nonetheless, when making inference about a node $X_{i}$, information about some other nodes beside its parents can reveal useful. In the limit case we could simply add all nodes in the graph but, among them, we know there might be some that are d-separated from $X_{i}$ and therefore, conditionally independent on $X_{i}$. For this purpose, a useful concept derived from the notion of d-separation and playing a central role in Bayesian Networks is that of \emph{Markov Blanket}. The Markov blanket of a node $X_{i}$ is the union of its parents, its children and its children's other parents and it corresponds to the set of nodes that makes all other nodes redundant when performing inference on a specific node (see Figure \ref{fig:mb}). It is worth noticing however, that not all nodes in the Markov Blanket of a node  are necessarily required to get complete probabilistic independence between $X_{i}$ and the nodes that are not in its Markov Blanket. Furthermore, the definition is symmetric: if a node $X_{j}$ is in the Markov Blanket of a node $X_{i}$, then the node $X_{i}$ is in the Markov Blanket of $X_{j}$.

The interpretation of the above mentioned measures of connection of a node to the whole network is straightforward. While the in-degree measures the number of nodes which have direct influence on a particular node (parent nodes), the out-degree measures the number of nodes that a particular node can directly influence (children nodes). On the other hand, the Markov blanket size of a node is always equal or greater than the sum of its in-degree and out-degree, since it is the sum of the number of its parent nodes, children nodes and the other parents of the children nodes. The Markov blanket size of a particular node is thus the number of nodes in the network that are involved in some conditional dependence relationship with it \citep{koller2009probabilistic}.

\subsubsection{Types of BNs}
\label{sec:type_of_BN}
Depending on the nature of the variables included in the model, we distinguish between:
\begin{enumerate}[label={\arabic*.}]
    \item \textbf{Discrete BNs}: all nodes are discrete variables
    \item \textbf{Continuous BNs}: all nodes represent continuous variables
    \item \textbf{Hybrid BNs}: nodes can represent either discrete or continuous variables.
\end{enumerate}
Moreover, looking at the temporal characterization of the phenomenon, it is possible to consider:
\begin{enumerate}[label={\arabic*.}]
\item \textbf{Static BNs}: the temporal dependence is not accounted for by the model and observations must be stochastically independent.
\item \textbf{Dynamic BNs}: the system evolves over time and the temporal dependence is accounted for by the model. The most common procedure to model a phenomenon dynamically by means of BNs requires to duplicate the original variables. That is, given $N$ variables considered at $S$ different lagged times, the resulting dynamical BN would have a total of $N\times S$ nodes to deal with \citep{ebert2012new}. Due to such an increase in the number of variables, the dynamic model commonly requires a bigger sample size, compared to the static case, in order to yield valid and reliable results.
\end{enumerate}

\subsubsection{Estimating a Bayesian Network}\label{dynamic}
estimating a Bayesian Network firstly requires to estimate the network structure (equivalent to the usual model selection problem) and secondly estimate the parameters of the global distribution.

\paragraph{Structure learning} 
In literature, algorithms for structure learning of a BN are generally classified into three different classes, depending on the adopted approach: constraint-based algorithms, score-based algorithms and hybrid algorithms \citep{zhou2011structure}. In this work we adopt a score-based approach by implementing the Hill-Climbing algorithm with $r=2$ random restarts \citep{scutari2014bayesian}. A score-based algorithm assigns each network a score (here the Bayesian Information Criterion (BIC)) and explores the DAGs' space in search for the structure that maximizes this score \citep{russell2016artificial}. 
 
Since the number of DAGs grows more than exponentially in the number of nodes, an exhaustive search in the space of all possible DAGs is barely feasible in practice. This is the reason as to why resorting to heuristic search algorithms, such as the Hill-Climbing, is very common in applications. More specifically, the Hill Climbing algorithm explores the DAGs space starting from a network, usually the one with no edges, and then adds, removes or reverses one edge at a time until the maximum gain in terms of the score is reached. Being the search strategy of a local type, there is no guarantee that the final structure recovered is indeed a global optimum. Therefore, in order to escape \emph{local optima}, further steps can be implemented such as \emph{random restarts} , that consist in randomly changing some arcs in the current optimal DAG, up to a number $r$ of times (see details in Algorithm \ref{algo:b}).

\begin{algorithm} \label{hill-climbin}
\DontPrintSemicolon 
\begin{enumerate}
  \item Start from a a graph $G$ (usually the empty graph)
  \item Compute $score_{G} = score(G)$
  \item Set $maxscore = score_{G}$
  \item Perform the following operations as long as $maxscore$ increases:
\begin{enumerate}
\item for each possible arc addition, deletion or reversal that results in a DAG
\begin{enumerate}
    \item Compute the score  $score(G^{*})$ of the modified structure $G^{*}$ \\and set $score_{G^{*}}= score(G^{*})$
    \item if $score_{G^{*}}> score_{G} $, set $G = G^{*}$ and  $score_{G} = score_{G^{*}}$
    \end{enumerate}
\item Update $maxscore = score_{G}$
\end{enumerate}
  \item  Up to $r$ times, perturb $G$ by adding, deleting and reversing multiple edges to obtain \\ a new DAG $G^{*}$ and search from step (IV).
  \item Return the DAG $G$.
\end{enumerate}

\caption{Hill Climbing algorithm with $r$ random restart \citep{scutari2014bayesian}}
\label{algo:b}
\end{algorithm}

\paragraph{Parameter learning}
After estimating the structure of a BN, it is possible to complete the description of the model by estimating the parameters of the global probability.

Depending on the variables nature, the three most common choices in literature, for which exact inference is indeed possible, are: Multinomial Bayesian Networks (all variables are discrete), Gaussian Bayesian Networks (all variables are continuous) and Conditional Linear Gaussian Bayesian Networks (CLGBNs) (variables can be either discrete or continuous) \citep{lauritzen1996graphical, scutari2012bnlearn}. Since we deal with both discrete and continuous variables, in this work we adopt Conditional Linear Gaussian Bayesian Networks. According to this model, the global probability distribution is assumed to be a mixture of multivariate Gaussian distributions with the constraint that continuous nodes cannot be parents of discrete nodes \citep{lauritzen1989graphical}. Additionally, the local distribution of a discrete node is a \emph{conditional probability table}, while the local distribution of a continuous node is a \emph{set of linear regression models}, one for each configuration of the discrete parents, with the continuous parents acting as regressors. Therefore, all probabilistic dependencies between continuous variables are assumed to be linear \citep{scutari2012bnlearn}.

\subsubsection{Encoding prior information in the learning algorithm}
\label{par:prior_info}
Both steps relating to structure and parameter estimation can be performed through numerical algorithms only (\emph{unsupervised learning}), or by combining algorithms that learn from the data with some sort of prior information on the phenomenon under investigation (\emph{supervised} learning).
In the context of structure learning, prior knowledge usually takes the form of specific edges that should or should not be present in the net \citep{scutari2012bnlearn}. The former case corresponds to providing the learning algorithm with a \emph{whitelist} that specifies the edges that must necessarily be present in the final DAG: they might be known from the literature or there might be a well established framework for the phenomenon under study that suggests the presence of specific edges. On the contrary, if prior knowledge is available in terms of connections that are not plausible, the learning algorithm is provided with a \emph{blacklist} that prevents some edges from being in the final DAG. This is the case, for instance, of  connections that are reputed as incoherent or impossible, or hard to be interpreted. As a direct consequence of including such prior information, the size of the space of DAGs is reduced, which generally helps to recover a better structure. The possibility of including prior information in terms of either a whitelist or a blacklist makes Bayesian Networks a very flexible class of models for real applications. 
   
 \subsubsection{Reliability of the estimated network }

In order to reduce model uncertainty derived from possible noise in the data and improve reliability of the learned structure, we adopt a \emph{model averaging approach} \citep{claeskens2008model, friedman1999learning, friedman1999data}. In place of estimating a single network structure from the data, a number $M$ of bootstrap datasets are created by resampling from the initial one and  $M$ (possibly different) network structures (i.e DAGs) are learned, one for each bootstrap dataset. From the $M$ learned structure, it is then possible to compute the relative frequency of each edge, called \emph{arc strength}, and choose only those edges that exceeds a fixed \emph{threshold}. The final network will thus comprise only those edges that are present in a "sufficiently large number" of estimated DAGs and, as a consequence, will generally be less sensitive to the noise in the data and perform better in prediction than the single models. A generally accepted "rule of thumb" for what the threshold should be suggests to keep only those edges that are present in at least $85\%$ of the learned DAGs, i.e. whose arc strength is larger than $0.85$. These arcs are then incorporated in the final network as directed edges with a specified direction if, conditioning on the edge's presence, this direction has a relative frequency bigger than $0.7$. Should this not be the case, the edge would be undirected \citep{scutari2014bayesian}. 

All analysis were done in R 2.14.0. The \emph{bnlearn} package  \citep{scutari2012bnlearn} was  used for Bayesian Network estimation, the \emph{bestNormalize} package  \citep{peterson2018bestnormalize} for variables' normalization and the \emph{bnstruct} package  \citep{franzin2016bnstruct} for handling missing values by means of the KNN algorithm.

\section{Results: applications of Bayesian Networks to the BES data}

 In this work, we implement Bayesian Networks for the BES atomic indicators in order to answer three main research questions:
       \begin{enumerate}
     \item What is the structure of the relationships, in terms of conditional independence statements, among the \emph{BES atomic indicators}? 
     \item What is the role that the different \emph{BES domains}, as they are currently defined and formalized in the BES framework, play in the whole BES system of indicators and how do they relate to each other?
     \item To what extent the structure of the relationships reflect the current BES theoretical framework? \end{enumerate}
     
For the purpose of satisfactorily answering the above inquiries, we deal with static, hybrid BNs that also incorporate geographic information and suggest two different strategies to encode prior information in the learning algorithm. The choice of static BNs was made in light of the small available time-window for the BES data ($8$ years) and the fact that this entirely falls into  the western economic crisis started in $2008$. The assumption of time homogeneity and the consequent choice of a static BN thus seemed to be reasonable, beside meeting the limitations of a small sample size (see section \ref{sec:type_of_BN}). 
Nonetheless, following, among others, \citep{vitolo2018modeling}, we still take into account - and partly model - some time dependency by considering the node YEAR as a discrete variable taking values in the set $\{2010,2011,2012,2013,2014,2015,2016,2017\}$. Hence, the choice of a hybrid BN. On the side, in order to deal with both discrete and continuous variables, the Conditional Linear Gaussian Bayesian Network model was implemented by means of the Hill-Climbing algorithm with $2$ random restarts and the BIC score. For the purpose of diminishing the noise present in the data and increase reliability of the estimated network, a model averaging approach was adopted, repeating the estimation procedure $1000$ times over $1000$ bootstrap samples from the original dataset and retaining only those edges that appeared in at least $85\%$ of the estimated structures. Additionally, we incorporate geographic information in the model  and, in order to consider the most suitable  geographic scale, four different geographical groupings of the Italian regions are contemplated and compared:

 \begin{description}
  \item[\textbf{Regional (NUTS 2 level)}] Geographic information is kept at the regional level and therefore the dataset consists of $63$ variables: $61$ continuous BES atomic indicators and $2$ discrete variables, namely YEAR and the spatial node REGION that takes up to $20$ different values (the number of the Italian regions). As  the BES database provided by  ISTAT is built at the regional level (see section \ref{second}), this  scenario corresponds to the finest  geographical partition that is possible to contemplate in the analysis.
  
  \item [\textbf{Macro-Areas (NUTS 1 level: NW-NE-C-S-I)}] In this scenario, BES atomic indicators are aggregated according to the NUTS 1 level macro-areas for the Italian peninsula. Therefore, the dataset consists of $63$ variables: $61$ continuous BES atomic indicators and $2$ discrete variables, namely YEAR and AREA, the last one taking values in \{North-West, North-East, Centre,  South, Islands\}.

  \item [\textbf{North/South}] Italian regions are grouped according to the dichotomy North/South. Thus, the dataset consists of $63$ variables: $61$ continuous BES atomic indicators and $2$ discrete variables, namely YEAR and AREA, the last one taking values in \{North, South\}.
  
  \item [\textbf{National}] According to this scenario, no relevant differences among regions  exist and no spatial label is added to the data, that therefore consists of a total of $62$ variables: $61$ continuous BES atomic indicators and the YEAR discrete variable. This scenario corresponds to a minimum of geographic information to include in the analysis.
 \end{description}

Finally, we did exploit the possibility of including prior information by formalizing blacklists and whitelists to pass to the learning algorithm (see Section \ref{par:prior_info}) and we suggest two different strategies aimed at analyzing relationships between the BES atomic indicators and relationships between the BES domains:
\paragraph{Strategy 1: blacklist only.} 
Prior knowledge takes the form of a blacklist of edges that are not allowed to be in the final network. Two groups of blacklisted edges are considered. The first group is made up of those edges that, in compliance with the restrictions from the CLGBN model, for which discrete node cannot be children of continuous nodes, are of the type $X\to Y$ where $X$ is continuous (i.e. $X$ is any BES atomic indicator) and $Y$ is discrete (i.e. $Y$ is either the variable YEAR or REGION/AREA). In the context of the present application, this assumption seems perfectly reasonable, since it represents the logical assumption that nor temporal or geographical variable can ever be influenced by the BES atomic indicators. In addition to this first group of blacklisted edges,  we also formalised a list of connections between BES atomic indicators considered as either unrealistic or incoherent and that, for this reason, should not be present in the final DAG. The complete list of the denied edges belonging to this second group can be found in Table \ref{blacklist} in Appendix.

\paragraph{Strategy 2: blacklist and whitelist.}
Both a blacklist and a whitelist are passed to the estimation algorithm. The blacklist is exactly the same as the one defined for Strategy $1$. On the other hand, the whitelist is built to reflect the hierarchical BES framework that relates the $61$ atomic indicators to $15$ different well-being domains and consists of all edges between pairs of atomic indicators belonging to the same domain. The usage of such a whitelist allows to investigate the relationships between the BES domains as they are currently formalized within the BES framework \citep{istat2018benessere}. In fact, from the point of view of the implemented score-based learning algorithm (Hill-Climbing with random restarts and BIC score), the introduction of a whitelist of edges between indicators in the same domain results in the algorithm choosing a smaller number of edges between indicators from different domains, if compared to the case of prior information consisting of a blacklist only. Clearly, this is ultimately due to the penalization mechanism brought about by the BIC used to score each network in the greedy search, which penalizes networks with a bigger number of parameters (that is, for a fixed number of nodes, networks with a bigger number of edges). As a consequence, a minor number of connections among different domains are expected, that will thus correspond to the strongest interactions in play, with the resulting network being able to provide a concise and clear picture of the interrelationships among well-being domains. As a final remark, it is worth  noticing that the usage of a whitelist defined in terms of edges between atomic indicators to inspect connections among domains can be considered as alternative to directly resorting to domain composite indicators, thus avoiding to potentially loose information along the synthesis process and dealing with variables whose meaning is maximally clear and transparent.

The two prior strategies should not be seen as competing, as they are designed to answer different questions. Specifically, Strategy  $1$ focuses on the relationships between atomic indicators and was implemented to answer the first research question, while Strategy $2$ aims at investigating connections among the different BES domains, as they are currently defined in the BES framework, and hence is devoted to answer the second question. These two first questions ultimately intend to scrutinize some of the mechanisms underlying the Italian regional well-being and therefore potentially give a contribution for the complex task of resources' allocation in policy decision making. Considerations derived from implementing both prior settings will then allow to answer the third research question, basically referring to the opportunity of highlighting some future directions for the BES system of indicators in light of the relationships arising from the network.

\subsection{Pre-processing}

The data pre-processing involved a step of  missing values imputation and a step  of variables' transformation to meet the 
Gaussianity assumption on the continuous variables required by the Conditional Linear Gaussian Bayesian Network (CLGBN) model. A total of  $103$  missing values occurred in the $160\times 61$ data matrix, (the $1\%$ of the whole dataset). Imputation of missing values was accomplished by K-Nearest Neighbour (KNN) algorithm \citep{duda2012pattern} and each missing value was substituted by the median of its $10$ nearest neighbours selected according to the Heterogeneous Euclidean-Overlap Metric (HEOM) distance \citep{wilson1997improved, franzin2016bnstruct}.\footnote{ According to this distance, given a set of continuous variable (or features) and two observations $x$ and $y$, the contribution of each feature to a measure of closeness between $x$ and $y$ is set to $1$ if the feature exhibits missing values in either $x$ or $y$ while, if values are not missing, it corresponds to their difference normalized by the feature range. To come up with a final distance between $x$ and $y$, all features' contributions are then aggregated by means of the euclidean distance.}

In order to meet the Gaussianity assumption, a set of transformations, including the logarithmic, arcsin, square root,  Box-Cox, Yeo-Johnson and the Ordered Quantile technique transformations were tested over each continuous variable, choosing the one that best performed according to the Pearson $P$ statistic for normality (see Tables \ref{gauss_1} and \ref{gauss_2} in Appendix) \citep{d1986goodness, peterson2019ordered, peterson2018bestnormalize}.

\subsection{Bayesian Network Estimation}
The structure learning algorithm was implemented for all four geographic scenarios (regional, macro-areas, North/South, national) with prior information encoded according to both Strategy $1$ and Strategy $2$, for a total of $8$ estimated networks. In order to adopt the most suitable geographical scale, a model choice was performed separately for each prior strategy by computing the Bayesian Information Criterion (BIC) and the Akaike Information criterion (AIC). In conformity to \citep{scutari2012bnlearn}, in this application AIC and BIC ($AIC= 2k-ln(\hat{L})$, $BIC = kln(n)-2ln(\hat{L})$), were rescaled by a factor -2, i.e. using $AIC = ln(\hat{L})/2- k$ and $BIC = ln(\hat{L})-kln(n)/2$,  where $\hat{L}$  is the maximized value of the likelihood
function of the model, $k$ is the number of parameters and $n$ the number of observations. Therefore larger
values are preferred over smaller values.

Additionally, predictive accuracy for each network was assessed through $k$-fold cross validation ($k=10$), by computing the average (among the continuous nodes) \emph{Posterior Mean Squared Error (MSE)} between the observed and the predicted values. The smaller this value, the higher the accuracy of the network. Table \ref{good} shows goodness-of-fit and predictive accuracy for each one of the eight estimated networks. The most appropriate configuration - among those tested - turned out to be the macro-areas subdivision (NUTS 1 level: NW-NE-C-SI) for both prior strategies. This geographic grouping outperformed the others in terms of both BIC and AIC while still resulting in good prediction accuracy in terms of the Posterior MSE. The estimated networks corresponding to this geographic scale and the two prior strategies are analyzed in the following paragraphs.

\subsubsection{Relationships among the BES atomic indicators: implementation of Strategy 1}

\begin{figure} 
  \centering
\includegraphics[width=19 cm,height=11.5 cm,keepaspectratio]{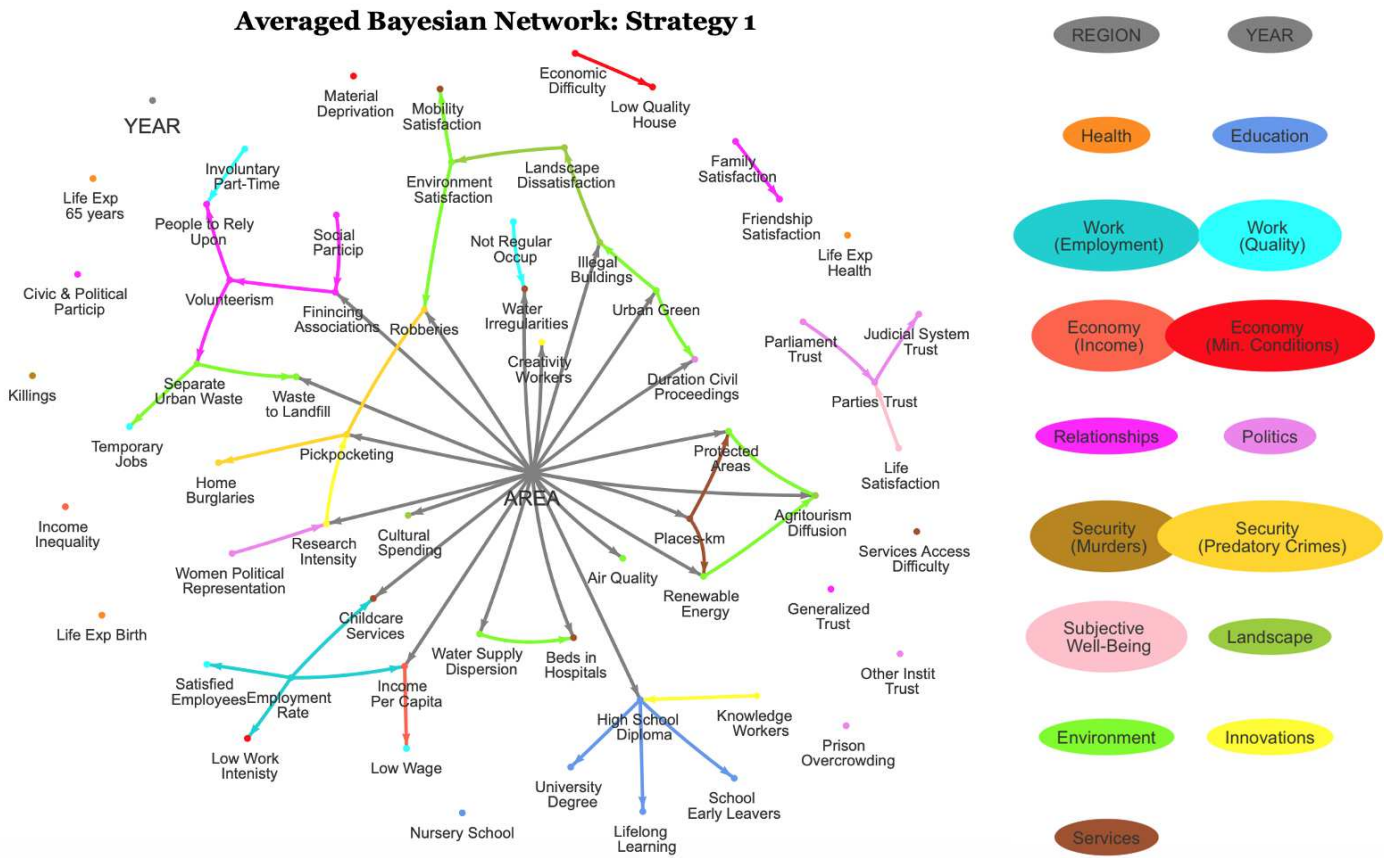}
\caption{Averaged network structure estimated by adopting prior Strategy 1 (blacklist only) with NUTS 1 level macro-areas (NW-NE-C-S-I).} 

  \label{fig:strategy_1}
\end{figure}

In order to answer the first research question about the structure of the relationships, in terms of conditional independence statements, among the BES atomic indicators, BNs were estimated by encoding prior information according to Strategy $1$. Figure \ref{fig:strategy_1} shows the network estimated in the prior setting induced by Strategy $1$ and corresponding to the Italian regions grouped into five macro-areas (NUTS 1 level: NW-NE-C-S-I). The analysis of the network will contemplate the network's topological configuration (analysis of the connected components) as well as the inspection of central nodes and particularly interesting structures among groups of nodes.

As far as the topology of the estimated network is concerned, it is possible to notice that, apart from $12$ isolated nodes (nodes that are not connected with any other node), the remaining of the $63$ nodes are grouped into $4$ major connected components, i.e. groups of nodes for which it is possible to go from one node to another one through edges that connect nodes in that group only. The larger connected component in this network is the one for which the geographic variable (AREA) is one of the root nodes (considering the nodes reordering in accordance to the arcs' direction). The relevance of considering the geographic position of the Italian regions, and specifically their grouping according to the NUTS 1 level, is revealed by the central role that the geographic variable has in this connected component. In fact, $21$ indicators pertaining to $10$ out of the $15$ BES domains have a \emph{direct connection} with the variable AREA while, if we also look at \emph{indirect connections}, i.e. connections that are mediated by some other node, the variable AREA turns out to influence $41$ out of the $61$ BES atomic indicators, for a total of $12$ affected BES domains. This result seems to be clearly interpreted as almost all features of well-being for the Italian regions showing differences that can  exclusively (direct connections) or partly (indirect connections) be explained by the geographic position. Furthermore,  almost all domains show to be connected among each other, in a way that intrinsically depends on which macro-area the regions belong to. This result is somehow expected, being very much consistent with the current inequality literature between, for example, northern and southern Italian regions, with southern regions generally suffering more from low employment rate and bad economic conditions \citep{checchi2010inequality,ferrara2019does, panzera2019measuring}. Interestingly, only three domains do not show a clear relationship with the geographic variable AREA, namely Health, Subjective Well-being and Security (Murders). Understanding the reason for this lack of connection is certainly of relevance within the well-being reasoning. A lack of influence of the variable AREA over these domains of well-being might indeed be motivated by these features being predominantly region-specific, with relevant differences even among regions belonging to same NUTS 1 level macro-area. Additionally, further investigation should also explore the possibility of some important variables being missing in the set of the BES atomic indicators currently used to build the estimated network, or even the opportunity for non-linear relationships that the current model (the CLGBN model) cannot recognize. 
The remaining three connected components formed by more than one node are smaller in size than the first component just considered and only refer to  one or maximum two different BES domains. In fact, the connected components $\{$\emph{Economic Difficulties, Low Quality House}$\}$ and  $\{$\emph{Friendship Satisfaction, Family Satisfaction}$\}$,  respectively refer to the Economic Well-being (Min. Conditions) and Relationship domains, while the component $\{$\emph{Parliament Trust, Parties Trust, Judicial System trust, Life Satisfaction}$\}$ includes variables from the Politics and Subjective Well-being domains.  Finally, the presence of nodes that are not connected to any other node should be interpreted as these variables being independent of any other variable in the network. These are therefore 
variables that are potentially able to bring an exclusive contribution to the picture of the Italian regional well-being,  by providing pieces of knowledge that are not related to the other considered aspects of well-being in terms of conditional independence relationships. This is the case for the set of indicators  \{\emph{Income Inequality, Material Deprivation, Nursery School, Services Access Difficulty, Prison Overcrowding, Civic \& Political Participation, Other Institution Trust, Killings}\}, and all  indicators in the Health domain, namely \{\emph{Life Expectancy at Birth, Life Expectancy in good health at Birth, Life Expectancy without activity limitations at 65 years of age}\}. The node YEAR is also an isolated node, meaning that no aspect of well-being is significantly influenced by the time variable, in the considered time window. This result supports our time-homogeneity assumption.

\begin{table}
 \caption{Goodness-of-fit and Predictive Accuracy of the estimated Bayesian Networks. In conformity to \citep{scutari2012bnlearn}, in this application AIC and BIC ($AIC= 2k-ln(\hat{L})$, $BIC = kln(n)-2ln(\hat{L})$), were rescaled by a factor -2, i.e. using $AIC = ln(\hat{L})/2- k$ and $BIC = ln(\hat{L})-kln(n)/2$,  where $\hat{L}$  is the maximized value of the likelihood
function of the model, $k$ is the number of parameters and $n$ the number of observations. Therefore larger
values are preferred over smaller values.}
 \label{good}

  \centering
  \begin{tabular}{cc|c|cc}
\\[-1.8ex]\hline 
\hline \\[-1.8ex] 
     & Spatial      & BIC & AIC  & Posterior    \\
     &  Scale  &  &   &  MSE   \\   
    
    \midrule
  
  & Regional & -11938 & -10843 & 0.239 \\
   Strategy 1    & \textbf{NW-NE-C-S-I} & \textbf{-10969} & \textbf{-10344} & 0.293 \\
 & N-S & -11380 & -11061 & 0.297 \\
  
  &National & -11433 & -11116 & 0.546\\

    \midrule
    
& Regional & -11503 & -10618 & 0.365 \\  
  Strategy 2 & \textbf{NW-NE-C-S-I} &  \textbf{-10238} & \textbf{-9206} & 0.383 \\
 & N-S & -10359 & -9680 & 0.391 \\

  &National & -10588 & -10041 & 0.380\\

\midrule
\end{tabular}
\end{table}

As a second step of the reading-through-the-network exercise, we look at particularly interesting structures and highlight the central role of some nodes in terms of their in-degree, out-degree and Markov blanket size (see Section \ref{MB}). These are measures of the degree of connection of a node to the remaining of the network and a complete description of all three features for each node is provided in Table \ref{degree} and Table \ref{degree2} in Appendix. All nodes have an in-degree varying from $0$ to $2$, and thus no node shows a central role in terms of the number of ingoing edges. On the other hand, three nodes show an interesting behaviour in terms of their out-degrees: AREA, \emph{Employment Rate} and \emph{High School Diploma}. We already highlighted the central role of the node AREA that, with an out-degree equal to $21$, is the node with the maximum number of outgoing edges, which clearly reflects the relevance of the geographic position of the Italian regions on the regional well-being as it is captured by the BES atomic indicators.
On the other hand, among the BES atomic indicators, \emph{Employment Rate} and \emph{High School Diploma} have the highest out-degrees, with $4$ and $3$ outgoing edges respectively, as well as the highest value of the Markov blanket size, which is equal to $5$ for both indicators. In order to better understand the role of these central indicators in the network, it is necessary to look at the structures in which they are inserted and, more specifically, the set of serial, diverging and converging connections (see Section \ref{fundamental_connections}) in which they are involved. 
In what follows, interesting structures (both including and not including  \emph{Employment Rate} and \emph{High School Diploma}) arising from the estimated network are analyzed, considering them in light of the engaged BES domains.

\paragraph{Work \& Life Balance, Economic Well-being and Services} \emph{Employment Rate} is connected to the remaining of the network by outgoing edges only and hence no BES atomic indicator influences it either directly nor indirectly. On the contrary, it affects $5$ indicators pertaining to $4$ different domains, namely \emph{Low Work Intensity} (Economic Well-being, [Min conditions]), \emph{Satisfied Employees} (Work \& Life Balance [Work Quality]), \emph{Childcare Services} (Services), \emph{Income per Capita} (Economic Well-being [Income \& Inequality]) and \emph{Low wage earners} (Work \& Life Balance [Work Quality]). Due to the diverging connections involved (which form something similar to a "cross" with \emph{Employment Rate} in the center), the first four indicators are pairwise independent given \emph{Employment Rate} or, in other words, it is the level of \emph{Employment Rate} that makes the four indicators pairwise marginally dependent. Therefore, any association detectable between, for example, \emph{Childcare Services} and \emph{Low Work Intensity} or \emph{Satisfied Employees} and \emph{Income per Capita} is due to the level of \emph{Employment Rate}. Another interesting structure involving \emph{Employment Rate} is the serial connection \emph{Employment Rate $\to$ Income per Capita $\to$ Low Wage}, for which \emph{Employment Rate} influences \emph{Low Wage} exclusively through \emph{Income per Capita} and, for this reason,  \emph{Employment Rate} and \emph{Low Wage} are independent if conditioning on \emph{Income per Capita}.  Finally, \emph{Employment Rate} also plays a role in two converging structures, namely \emph{Employment Rate $\to$ Income per Capita $\leftarrow$ AREA} and \emph{Employment Rate $\to$ Childcare Services $\leftarrow$ AREA}. From these two connections it is possible to scrutinize the dependence of \emph{Employment Rate} from AREA by assessing that  \emph{Employment Rate} and AREA are marginally independent and become dependent when conditioning on either the level of \emph{Income per Capita} or \emph{Childcare Services}. In other words, the association between \emph{Employment Rate} and AREA can be explained by either considering the economic wealth (as it is measured by \emph{Income per Capita}) or the availability of municipal childcare services (measured by \emph{Childcare Services}). The central role of \emph{Employment Rate} in influencing variables related to work conditions, work satisfaction, income and childcare services, partly reflects results in, for example, \citep{maitah2015exploring, kitov2011employment, kreyenfeld2000does} and can be potentially relevant for decision making policies. 
Finally, we report on the indicators  \emph{Economic Difficulty} and \emph{Low Quality House} from the Economic Well-being (Min. Conditions) domain, which are not involved in this group of connections but form a connected component on their own, with \emph{Low Quality House} directly reflecting the level  of \emph{Economic Difficulty} and the latter being marginally independent on any other node.

\paragraph{Education and  Research}
The node \emph{High School Diploma} from the Education domain shows both ingoing and outgoing edges. It directly influences $3$ indicators pertaining to the Education domain, namely \emph{University Degree}, \emph{Lifelong Learning} and \emph{School Early Leavers} and it is directly affected by the nodes AREA and \emph{Knowledge Workers} from the Innovation domain. Due to the serial connections involved in this group of nodes (which overall forms something similar to a "star" with \emph{High School Diploma} in the center), the three indicators from the Education domain and the node AREA are marginally dependent and become (conditionally) independent when conditioning on the level of \emph{High School Diploma}. Therefore, it is indeed the level of \emph{High School Diploma} that explains the relationship between AREA and  \emph{University Degree}, \emph{Lifelong Learning} and \emph{School Early Leavers}. Similar reasoning holds when AREA is substituted by \emph{Knowledge Workers}: the dependence between the latter and the three indicators \emph{University Degree}, \emph{Lifelong Learning} and \emph{School Early Leavers} is explained by the level  of \emph{High School Diploma}. Therefore, \emph{High School Diploma} mediates the effect of both AREA and \emph{Knowledge Workers} on the other Education indicators. Finally, \emph{High School Diploma} is also involved in the  converging structure AREA $\to$ \emph{High School Diploma} $\leftarrow$ \emph{Knowledge Workers}, from which it is possible to assess that AREA and \emph{Knowledge Workers} are marginally independent and become conditionally dependent if the level of \emph{High School Diploma} is known.  As far as the Education domain is concerned, it is noteworthy that \emph{Nursery School} is the only indicator among the ones used to build the composite indicator for Education (see Section \ref{second} and Table \ref{tab:bes1} in Appendix), that does not join this ("star") structure. In fact, it is an isolated node in the estimated network (constituting a connected component on its own) and therefore  is (marginally) independent on any other node. This result is somehow reasonable, since the four connected indicators all concern education for the adults (people aged $18$ or above), while \emph{Nursery School} is the only indicator in the Education domain measuring the phenomenon at the children level (people aged between $4$ and $5$). The estimated network confirms the lack of any association between the two groups of indicators representing adult education and children education and could eventually lead to a further discussion on the extent to which children education is properly represented in the final composite indicator for Education or even question the suitability of keeping adult and children education indicators together in the same composite indicator. 

\paragraph{Subjective Well-being and Politics}
Another interesting structure worth analyzing is the one arising from the connected component $\{$\emph{Parliament Trust, Parties Trust, Judicial System trust, Life Satisfaction}$\}$. Here two aspects emerge. First of all, Subjective Well-being domain is connected to three over the four indicators in the Politics domain that measure citizens' trust in political institutions. More specifically,  \emph{Life Satisfaction} directly affects \emph{Parties Trust} and is indirectly connected to \emph{Parliament Trust} and \emph{Judicial System Trust}. Secondly, \emph{Parties Trust} seems to have a central role in this relationship, as it mediates for the influence of \emph{Life Satisfaction} on \emph{Judicial System Trust} (due to the corresponding serial connections) and makes \emph{Parliament Trust} and \emph{Life Satisfaction} conditionally dependent (due to the corresponding converging structure). Subjective Well-being affecting Politics is clearly an interesting feature of well-being and is consistent with results in \citep{ceriani2016multidimensional} where the same is assessed in the context of a study of well-being in Eastern and Western Europe by means of data from the Life in Transition Survey (LITS II).

\paragraph{Environment, Landscape, Services and Security (Predatory Crimes)}
Indicators from the domains of Environment, Landscape, Security and Services clearly appear to be interrelated with each others in more than one part of the network. First of all,
indicators related to the satisfaction for the living place, meant as related to either the environment, landscape or mobility, are connected to each other. This is expressed by the serial connection  \emph{Landscape Dissatisfaction} $\to$ \emph{Environment Satisfaction} $\to$ \emph{Mobility Satisfaction}, where \emph{Environment Satisfaction} mediates the relationship between the other two. Furthermore,  \emph{Landscape Dissatisfaction} is also involved in the serial connection \emph{Urban Green} $\to$ \emph{Illegal Buildings} $\to$ \emph{Landscape Dissatisfaction}, with  \emph{Illegal Buildings} explaining the relationship between \emph{Urban Green} and \emph{Landscape Dissatisfaction}, and both \emph{Illegal Buildings} and \emph{Urban Green} being directly influenced by AREA. In the second place, aspects of well being related to the living place satisfaction  are interestingly connected to all indicators in the  Security (Predatory Crime) domain. In fact, \emph{Environment Satisfaction} directly influences \emph{Robberies} (which is also directly affected by the node AREA) and indirectly influences all other indicators of predatory crimes as it is described by the following concatenation of serial connections \emph{Environment Satisfaction} $\to$  \emph{Robberies}, $\to$ \emph{Pickpocketing} $\to$ \emph{Home Burglaries}. Finally, the Environment, Landscape and Services domains also interact with each other in that some indicators from the Services domain appear to be connected to indicators devoted to measuring environmental sustainability. This is the case, for example, for the diverging connection \emph{Protected Areas} $\leftarrow$ \emph{Places-Km} $\to$ \emph{Renewable Energy} where two environmental indicators are made independent if conditioned on an indicator from the Services domain, and for the serial connection \emph{Water Supply Dispersion} $\to$ \emph{Beds in Hospital} where \emph{Beds in Hospital} reflects what is measured by \emph{Water Supply Dispersion}, both indicators being also strongly influenced by the geographic position of the region.

\paragraph{Environment, Work \& Life Balance and Relationships}
Interestingly, indicators related to having a stable work, in terms of either the number of hours worked in a week (\emph{Involuntary Part-Time}) or the type of contract (\emph{Temporary Jobs}), connects to aspects of social participation, as they are described by the Relationships indicators  \{\emph{Volunteering, Social Participation, Financing of Associations, People to Rely Upon}\}, and to indicators from the Environment domain that pertain to the waste management process, namely \{\emph{Separate Urban Waste, Waste to Landfill}\}. This overall group of nodes could potentially be seen as representing those traits of well-being strictly related to the possibility for the citizens to have an active role in their community. In fact, this opportunity certainly reflects into the undertaken social activities (e.g. volunteering, social participation, financing association...), but also the job contractual conditions and, perhaps more surprisingly, the behaviour related to separating urban waste which, in turn, also directly influences the transfer of urban waste to landfill. Interestingly, \emph{Family Satisfaction} and \emph{Friendship satisfaction} from the Relationship domain are not involved in this group of connections, forming a connected component on its own.

\subsubsection{Relationships among the BES domains: implementation of Strategy 2}
\begin{figure} 
  \centering
\includegraphics[width=19 cm,height=11.5 cm,keepaspectratio]{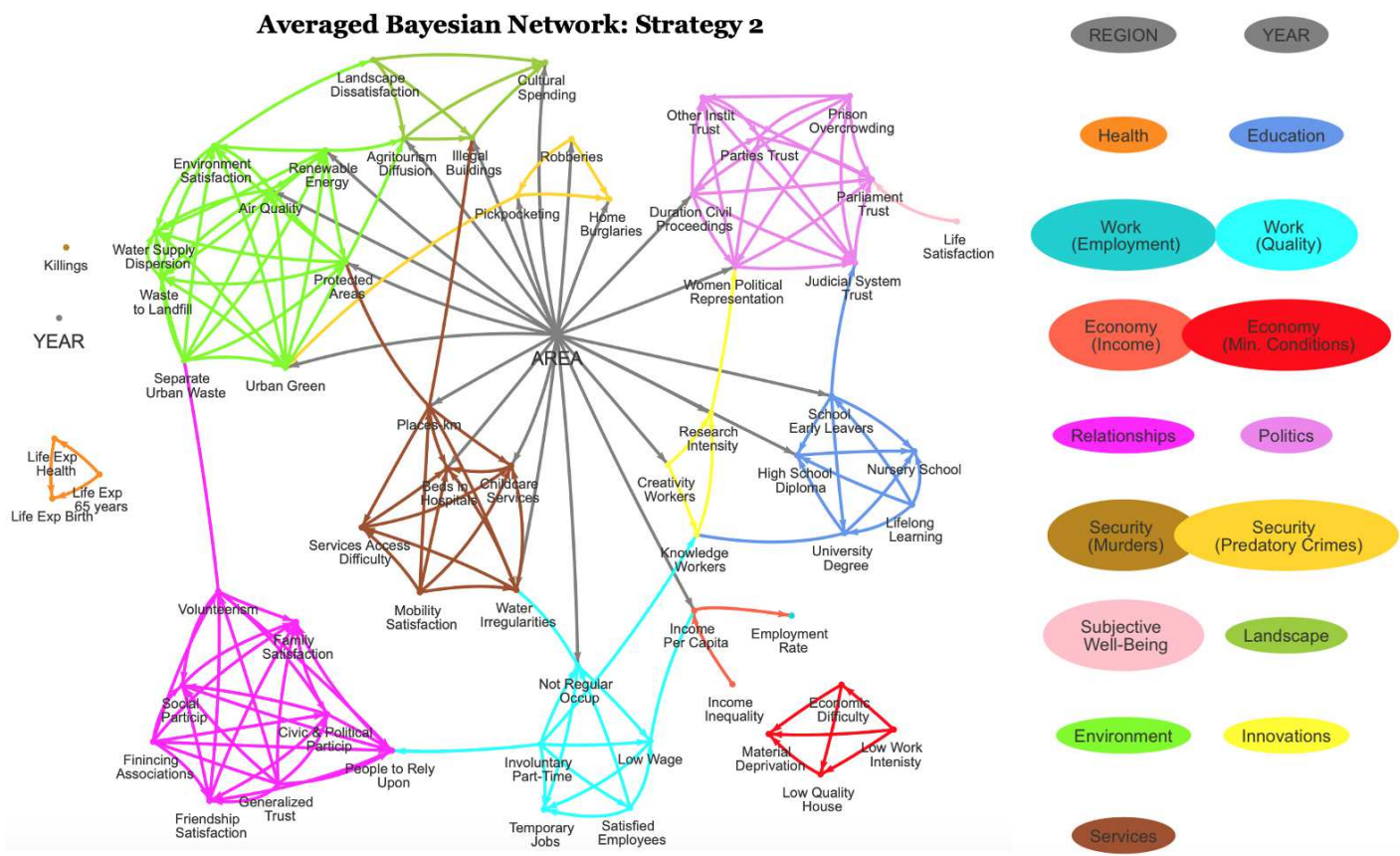}
\caption{Averaged network structure estimated by adopting prior Strategy 2 (blacklist and whitelist) with NUTS 1 level macro-areas (NW-NE-C-S-I).}
  \label{fig:model2}
\end{figure}

In order to answer the second research question concerning the relationships and roles of the different BES domains as they are currently defined within the BES framework, BNs were estimated by encoding prior information according to Strategy 2. The only difference with the Strategy 1 set up is that atomic indicators pertaining to the same domain are here compelled to be connected. Figure \ref{fig:model2} shows the structure of the network estimated using the prior setting induced by Strategy $2$ and corresponding to the Italian regions grouped into five macro-areas (NUTS 1 level: NW-NE-C-S-I). As for Strategy $1$, the analysis will focus on the structure of the network and will proceed by considering the network's topological configuration (analysis of the connected components), as well as the study of BES domains showing a central role and particularly interesting connections among domains.

In terms of its topological structure, the network is composed of $5$ connected components. The largest component includes the node AREA as well as $12$ out of the $15$ BES domains, with the only exception of Health, Economic Well-being (Min. Conditions) and Security (Murders). This shows, as expected, a high level of interaction among the BES domains, as well as the central role of the regional geographic position in intervening on these interactions, with the node AREA directly influencing $9$  BES domains. Three out of the four remaining connected components correspond to the domains Health, Economic Well-being (Min. Conditions) and Security (Murders) which do not show specific connections with the other domains. For the case of Economic Well-being (Min. Conditions) and Security (Murders), this configuration agrees with the recent development within the formalization of the BES domains, corresponding to the splitting of the Economic Well-being and Security into their sub-components \citep{istat2018benessere}.   The last connected component is uniquely formed by the YEAR node, confirming the lack of significant influences of the temporal variable over the indicators' relationships also within this prior setting.

For the purpose of quantifying the degree of connection of each domain with the remaining of the network, and consequently identify possible domains having a particularly central role, we resort to the number of directly connected components of a domain, i.e. the number of domains that have at least one direct connection with it. The value of this summery statistics for each domain is reported in Table \ref{strategy2} in Appendix. All domains are directly connected with a number of domains between $0$ to $4$ and the most connected domains are Work \& Life Balance (Work Quality) and Environment, with $4$ directly connected domains each. Specifically, the Environment domain is directly connected to the domains Landscape (by means of indicators related to the living place satisfaction), Security (Predatory Crimes) (by means of the presence of \emph{Urban Areas}), Services (through the indicator \emph{Places-km}) and Relationships (thanks to the connection between \emph{Separate Urban Waste} and \emph{Volunteering}). On the other hand, the Work \& Life Balance (Work Quality) domain has direct connections with the domains Economic Well-being (Income) (by means of the edge \emph{Low Wage} $\to$ \emph{Income per Capita}), Innovation (through the indicator \emph{Knowledge Workers}), Relationships  and Services. The high number of direct connections for the Environment and Work \& Life Balance (Work Quality) domains suggests a central role for them in the whole BES system, in that they are potentially able to influence (or at least reflect) several other traits of well-being. 
In addition to relationships that concern the highest connected domains, there is a number of connections that are worth highlighting and that are, to a certain extent, in agreement to what observed from the network estimated in the prior setting of Strategy $1$. First of all, Subjective Well-being confirms to be able to directly influence (or at least reflect) the Politics domain. This connection is realized through the edge \emph{Life Satisfaction} $\to$ \emph{Parliament Trust}, which seems to confirm that the existing relationships between the two domains is indeed due to the subgroups of indicators in the Politics domain that refer to the level of trust of citizens towards political institutions. The Politics domain is also directly connected to the Innovation and Education domains by means of the edges \emph{ Research Intensity $\to$ Women Political Representation} and \emph{School Early Leavers $\to$ Judicial System Trust}, respectively. Moreover, strict interactions emerge between aspects related to work and economic wealth, with Economic Well-being (Income) mediating the relationship between the two domains of Work \& Life Balance (Work Quality) and Work \& Life Balance (Employment). Finally, it seems interesting to notice the direct connection among the domains Education, Innovation and Work \& Life Balance (Work Quality) by means of the node \emph{Knowledge Workers}.

\subsubsection{Consistencies and novelties with respect to the BES system of indicators: combining results from the two prior strategies}
Considerations derived from implementing both prior settings will allow to answer the third research question, investigating the extent to which the structure of the estimated relationships reflect the current BES theoretical framework. This purpose basically refers to the opportunity of highlighting some future directions for the BES system of indicators in light of the relationships arising from the networks. 

From the network estimated by implementing Strategy $1$, we can assess that the BES theoretical framework consisting of domains and corresponding atomic indicators is indeed reflected in the estimated network. In fact, the network could recover (at least some) parts of almost all BES domains, by connecting indicators belonging to the same domain for $10$ out of the $15$ BES domains. This is particularly evident for the domains Education, Economic Well-being (Min. Conditions), Work \& Life Balance (Quality), Landscape, Environment, Relationships, Politics and Security (Predatory Crimes). It is noteworthy that the network could even reflect the sub-domains that have recently been introduced by ISTAT  with the intent of splitting the domains of Economic Well-being, Work \& Life Balance and Security into two sub-components each \citep{istat2018benessere}. In fact, indicators in the Economic Well-being (Min. conditions) and Economic Well-being (Income) domains never connect with each other, while indicators in the same sub-component do show associations. The same holds for Work \& Life Balance (Work Quality) and Work \& Life Balance (Employment) and for Security (Predatory Crimes) and Security (Murders). 
 
Identifying the right domains as well as choosing the atomic indicators informing those domains, is a crucial exercise in order to get a meaningful measurement of the phenomenon of interest. Particularly, this is the key instrument through which balancing a detailed description of the phenomenon -realized by looking, for instance, at hundreds of indicators- with the need of an efficient synthesis of the overall available information. This is precisely the perspective from which reading the following considerations arising from the analysis of the network. In fact, the BN analysis seems to suggest some other possible splittings in light of the different roles that groups of indicators, currently referred to the same BES domain, seem to have in the network’s functioning. An example is given by the Environment domain, with reference to which it is possible to identify the sub-component \{\emph{Separate Urban Waste, Waste to Landfill}\}, to be conceivably ascribed to the Waste Management Process. A second example is in the Relationships domain, for which two sub-components can potentially be identified, namely the group \{\emph{Volunteering, Social Participation, Financing of Associations, People to Rely Upon}\}, related to aspects of social participation and the group \{\emph{Friendship Satisfaction, Family Satisfaction}\}, concerned with the satisfaction for social relationships, the latter sub-component even forming a connected component on its own. Finally, some considerations can also be raised about the Education domain, where the set of indicators \{\emph{University Degree}, \emph{Lifelong Learning} and \emph{School Early Leavers}\} specifically refers to aspects of the adults education and formation, while \emph{Nursery School} is the only indicator (among those used to build the composite indicator for the domain Education) that is currently in charge of measuring children's education.

Finally, by looking at the network estimated as in the setting of Strategy $2$, we note that almost all domains are connected to each other by means of at most one direct edge, which can be interpreted as the BES domains, which are expected to catch diverse, although related, aspects of well-being, being distinguished from each other to a good extent. The only exception being the Environment and Landscape domains, which show a higher level of interrelation being directly connected through three edges, which could lead to perceive these two domains as not so clearly distinguishable. 

\section{Conclusions and Future Remarks}

This work aims to be a first attempt to approach the complex phenomenon of multidimensional well-being in the Italian regions through the tool of the Bayesian Networks' class of models. The possibility of combining probabilistic assessments with a graphical representation seemed to be promising for a dataset with a considerable number of variables that poses the question of how to fruitfully deal with the amount of uncertainty present in it. Specifically, BNs can considerably simplify the complexity of the relationships among different aspects of a phenomenon by highlighting the most significant connections and build an overall model of interactions by integrating data and prior knowledge.

We showed that Bayesian Networks can be considered as a suitable class of models for studying the multidimensional well-being for the Italian regions, being also able to incorporate geographic information. We suggested two different settings for encoding prior information in the network learning algorithm, which allow to conduct the analysis at two different levels of the hierarchical framework of the BES, by either looking at the relationships among the BES atomic indicators or relationships among the BES domains. Considerations arising from both levels of the analysis could finally lead to formulate some considerations for possible future development of the BES system of indicators.

We believe that this work can develop in a number of possible, future directions. By way of illustration, different learning algorithms can be considered for the Conditional Linear Gaussian Bayesian Networks (CLGBN) model, as well as alternative models that could, for example, account for possible dependencies among regions (which were not contemplated in this work). Moreover, gender differences can also be potentially inspected by comparing gender-specific network structures and the availability of data referring to a larger number of years could allow to include other variables in the analysis, as for example those BES indicators that are currently not used to compute composite indicators for each domain. Finally, other directions for future works could also include the attempt to integrate the results from the estimated network in the development of a model-based classifier to characterize, and possibly rank, the Italian regions according to their well-being.

\appendix
\section*{Appendix} \label{app:a}
\begin{table}
 \caption{BES dataset: BES domains and corresponding atomic indicators (Outcome domains) \citep{istat2018benessere}}
  \scriptsize 
  \centering
  \begin{tabular}{llll}
\\[-1.8ex]\hline 
\hline \\[-1.8ex] 
      Domain (Label)     & Indicator (Label)  & Polarity \\
    
    \midrule
      Health & Life expectancy at birth (Life Exp Birth)   & + \\
    & Life expectancy in good health at birth (Life Exp Health)  & +\\
    &  Life expectancy without activity limitations at $65$ years of age (Life Exp 65 years)  & +\\ 
    
    \midrule
     Education    & Participation in nursery school (Nursery School) & + \\

     & People with at least a high school diploma (High Scool Diploma) & +\\
     & People who have 
university degree or have completed tertiary (University Degree) & +\\ 
 &  \qquad \quad  \ education ($30-34$ years old) \\
      & Early leavers from education and training (School Early Leavers) & -\\
     & Participation in lifelong learning ($25-64$ years old)  (Lifelong Learning)& + \\
     
     \midrule
     Work \& Life Balance & Employment rate  & +\\
   \scriptsize{EMPLOYMENT} & & \\
\midrule
 Work \& Life Balance & Employed persons with temporary jobs at least for $5$ years (Temporary Jobs) & -\\
  \scriptsize{WORK QUALITY} & Low wage earners (Low Wage) & -\\
& Employed persons not in regular occupation (Not Regular Occup)  & -\\
& Employed persons who feels satisfied with their works (Satisfied Employees)  & +\\
& Involuntary Part-time employed persons (Involuntary Part-Time)   & -\\
\midrule

    Economic Well-being & Income per Capita (Income per Capita) \\
    \scriptsize{INCOME AND INEQUALITY} & Disposable income inequality (Income Inequality)  & + \\
   \midrule

    Economic Well-being & Serious material deprivation (Material Deprivation)  & - \\
     \scriptsize{MINIMUM ECONOMIC}  & Low quality of the house (Low Quality House)  & -\\
   \scriptsize{CONDITIONS}& Great economic difficulty (Economic Difficulty)  & -\\
   & Very low work intensity (Low Work Intensity)  & -\\
   \midrule

Social Relationships & Satisfaction with family relationships (Family Satisfaction)  & +\\
(Relationships)    & Satisfaction with friendships  (Friendship Satisfaction)& +\\
    & People to rely upon (People to Rely Upon)  & +\\
    & Social participation (Social Particip)  & + \\
    & Civic and political participation (Civic and Political Particip) & +\\
     & Voluntary activities (Volunteerism)  & +\\
     & Financing of associations (Financing Associations)  & +\\
     & Generalized trust (Generalized Trust)  & +\\
    \midrule
    
    Security & Index of killings (Killings)  & -\\
     \scriptsize{MURDERS} \\
        \midrule
       Security & Home burglaries  (Home Burglaries)& -\\
     \scriptsize{PREDATORY CRIMES}  & Pickpocketing (Pickpocketing)  & -\\
    & Robberies (Robberies)  & -\\
    
    \midrule
    Subjective Well-being & Overall life satisfaction (Life Satisfaction)  & +\\
    \midrule

    Landscape \&  &  Current spending for
heritage management by local administrations (Cultural Spending)  & +\\
Cultural Heritage& Illegal building (Illegal Buildings)  & - \\
(Landscape) & Diffusion of agritourism companies (Agritourism Diffusion)  & +\\
&  Dissatisfaction with the landscape of the place of life (Landscape Dissatisfaction)  & -\\

        \midrule
    Environment & Dispersion from the municipal water supply (Water Supply Dispersion)  & - \\
    & Transfer of urban waste to landfill (Urban Waste to Landfill)  & +\\
    & Air quality (Air Quality)\\
    & Urban green availability (Urban Green)  & +\\
    & Satisfaction with the environmental situation (Environment Satisfaction)  & +\\
    & Protected land areas (Protected Areas)  & +\\
    & Energy from renewable sources (Renewable Energy)  & +\\
    & Separate collection of urban waste (Separate Urban Waste)  & +\\
    
    \midrule
  \end{tabular}
  \label{tab:bes1}
\end{table}

\begin{table}
 \caption{BES dataset: BES domains and corresponding atomic indicators (Context domains)}
 \scriptsize 
  \centering
  \begin{tabular}{lllc}
\\[-1.8ex]\hline 
\hline \\[-1.8ex] 
      Domain (Label)     & Indicators (Label)  & Polarity  \\

    \midrule
     Politics \& Institutions & Trust in the Italian parliament (Parliament Trust) & + \\
    (Politics) & Trust in the judicial system (Judicial System Trust)  & +\\
    & Trust in parties (Parties Trust)  & +\\
    & Trust in other types of institutions (Other Instit Trust) & +\\
    & Women and political representation at local level (Women Political Representation)  & +\\
    & Duration of civil proceedings (Duration Civil Proceedings)  & -\\
    & Prison overcrowding (Prison Overcrowding)  & -\\
    
    \midrule
      Research, Innovation  & Research intensity (Research Intensity)  & +\\
     \& Creativity& Knowledge workers (Knowledge Workers ) & +\\
     (Research)& Persons employed in creative enterprises (Creativity Workers)  & +\\
       \midrule
     
    Quality of Services & Beds in residential social-welfare and social-health facilities (Beds in Hospital)  & +\\
   (Services) & Children who have benefited from municipal childcare services (Childcare Services)  & +\\
    & Difficulty in accessing some service (Services Access Difficulty)s  & -\\
    & Irregularity in water distribution  (Water Irregularities)& -\\
    & Places-km offered by local public transport (Places-km)  & +\\
    & Satisfaction with mobility services  (Mobility Satisfaction)& +\\
    \bottomrule
  \end{tabular}
  \label{tab:bes2}
\end{table}

\begin{table}[!htbp] \centering 
  \caption{
  Pearson P test statistic for Gaussianity and the corresponding used transformation for each atomic indicator (Outcome domains).} 
  \label{gauss_1} 
  \scriptsize 
\begin{tabular}{@{\extracolsep{5pt}} llll} 
\\[-1.8ex]\hline 
\hline \\[-1.8ex] 
 Domain& Indicator & $P/dp$ & Transformation \\ 
\hline \\[-1.8ex] 
 Health& Life Exp Birth & $1.3$ & Yeo-Johnson \\ 
 & Life Exp Health & $1.4$ & Ordered Quantile  \\ 
 & Life Exp 65 years & $1.0$ & Ordered Quantile \\ 
  \midrule
 Education  & Nursery School & $1.3$ & Box-Cox \\ 
 & High School Diploma & $0.7$ & Ordered Quantile \\ 
 & University Degree& $1.1$ & no transform \\ 
 & School Early Leavers & $1.1$ & Box-Cox  \\ 
 & Lifelong Learning & $0.5$ & no transform \\ 
\midrule
  Work \& Life Balance   & Employment Rate & $1.5$ & Ordered Quantile \\ 
  \scriptsize{EMPLOYMENT}&&\\
  \midrule
  Work \& Life Balance   & Temporary Jobs & $0.9$ & Box-Cox  \\ 
   \scriptsize{WORK QUALITY}& Low Wage& $1.1$ & Ordered Quantile \\ 
 & Not Regular Occup& $1.2$ & Ordered Quantile \\ 
 & Satisfied Employees & $1.8$ & Ordered Quantile \\ 
 & Involuntary Part-Time& $1.1$ & no transform \\ 
  \midrule
   Economic Well-being& Income Per Capita& $0$ & arcsin \\ 
\scriptsize{INCOME AND} & Income Inequality & $0.7$ & Ordered Quantile \\ 
 \scriptsize{INEQUALITY} &&\\
 \midrule
 Economic Well-being  & Material Deprivation & $0.8$ & Ordered Quantile \\ 
\scriptsize{MINIMUM ECONOMIC} & Low Quality House & $0.8$ & Box-Cox  \\ 
\scriptsize{CONDITIONS} & Economic Difficulty& $0.9$ & arcsin \\ 
 & Low Work Intensity & $1.4$ & log\\ 
  \midrule
Relationships   & Family Satisfaction & $1.1$ & Square Root \\ 
 & Friendship Satisfaction & $0.9$ & no transform \\ 
 & People to Rely Upon & $1.0$ & Yeo-Johnson \\ 
 & Social Participation & $1.0$ & no transform \\ 
 & Civic and Political Participation & $1.2$ & Ordered Quantile \\ 
 & Volunteering & $0.8$ & log \\ 
 & Financing Associations& $1.4$ & no transform \\ 
  & Generalized Trust& $1$ & Square Root \\ 
 \midrule
 Security & Killings & $0$ & arcsin \\ 
 \scriptsize{MURDERS} &&\\
 \midrule
 Security & Home Burglaries & $1.2$ & Ordered Quantile \\ 
  \scriptsize{PREDATORY CRIMES}& Pickpocketing & $0.9$ & Ordered Quantile \\ 
& Robberies & $2.1$ & Square Root \\ 
 \midrule
 Subjective Well-being & Life Satisfaction & $1.5$ & no transform \\ \midrule
 Landscape& Cultural Spending & $1.5$ & Ordered Quantile \\ 
 & Illegal Buildings & $1.1$ & Ordered Quantile \\ 
 & Agritourism Diffusion & $0.6$ & arcsin \\ 
 & Landscape Dissatisfaction & $0.7$ & log \\ 
 \midrule
  Environment& Water Supply Dispersion & $1.6$ & arcsin \\ 
 & Waste to Landfill & $1.1$ & Yeo-Johnson \\ 
 & Air Quality & $1.2$ & Ordered Quantile \\ 
 & Urban Green & $1.1$ & Ordered Quantile \\ 
 & Environment Satisfaction & $1.2$ & Yeo-Johnson \\ 
 & Protected Areas & $1.4$ & Ordered Quantile \\ 
 & Renewable Energy & $1.0$ & arcsin \\  & Separate Urban Waste & $0.7$ & Ordered Quantile \\ 

\hline \\[-1.8ex] 
\end{tabular} 
\end{table}

\begin{table}[!htbp] \centering 
  \caption{Pearson P test statistic for Gaussianity and the corresponding used transformation for each atomic indicator (Context domains).}   \label{gauss_2} 
\scriptsize
\begin{tabular}{@{\extracolsep{5pt}} llll} 

\\[-1.8ex]\hline 
\hline \\[-1.8ex] 
 Domain& Indicator & $P/dp$ & Transformation \\ 
\hline \\[-1.8ex] 

 Politics& Parliament Trust& $1.6$ & Box-Cox \\ & Judicial System Trust & $1.1$ & Box-Cox \\ 
 & Parties Trust & $1.2$ & arcsinh \\ 
 & Other Institutions Trust & $2.6$ & Box-Cox \\ 
 & Women Political Representation & $0$ & arcsin\\ 
 & Duration Civil Proceedings& $0$ & arcsin \\ 
 & Prison Overcrowding & $0.7$ & no transform\\ 
 \midrule
Research & Research Intensity & $1.1$ & log \\ 
 & Knowledge Workers& $0.9$ & no transform \\ 
 & Creativity Workers & $0.7$ & log\\ 
 \midrule
 Services & Beds in Hospitals & $1.1$ & Ordered Quantile \\ 
 & Childcare Services & $1.3$ & no transform \\ 
 & Services Access Difficulty & $1.6$ & Ordered Quantile \\ 
 & Water Irregularities & $0.6$ & Ordered Quantile \\ 
 & Places-km & $0$ & arcsin \\ 
 & Mobility Satisfaction & $1$ & Ordered Quantile\\ 
\bottomrule
\end{tabular} 
\end{table}

\begin{table}
 \caption{Prior blacklist: list of the edges $X \to Y$ not allowed to be present in the network structure.}
 \scriptsize
  \centering
  \begin{tabular}{ll}
\\[-1.8ex]\hline 
\hline \\[-1.8ex] X & Y\\
\midrule
Life expectancy at birth & Air quality\\
Life expectancy at birth & Protected Areas\\
Pickpocketing & Intensity of Research\\
Pickpocketing & Women and political representation at local level\\
Home Burglaries & Proteced Areas \\ 
Robbery & Women and political representation at local level\\
Overcrowding & People to rely Upon\\
Overcrowding & Participation in lifelong learning\\
Overcrowding &  Women and political representation at local level\\
Overcrowding & Involuntary Part-time employed persons \\
Overcrowding & Protected Areas\\
Abusivism & Intensity of Research\\
Places km & Intensity of Research\\
Water dispersion & Women and political representation at local level\\
Water dispersion & Intensity of Research\\
Air Quality & Participation in lifelong learning\\
Generalized Trust & protected Areas\\
Trust in Judicial System & Life expectancy without activity limitations at65years of age\\
Trust in Parliament & Air Quality \\
Civil Proceedings & Beds in residential social-welfare and social-health facilities\\
Social Participation & Protected Areas\\
Civic and Political Participation & Serious material deprivation \\
Participation in nursery school & Protected Areas\\
Separate collection of urban waste & Women and political representation at local level\\
Children who have benefited from municipal childcare services & Protected Areas\\
\bottomrule
 \label{blacklist}
\end{tabular}
\end{table}

\begin{table}[!htbp] \centering 
  \caption{Network estimated by adopting prior Strategy 1 with NUTS 1 level areas (NW-NE-C-S-I): in-degree, out-degree and Markov blanket size of the nodes (Outcome domains).} 
  \scriptsize
  \label{degree} 
\begin{tabular}{@{\extracolsep{5pt}} lllll} 
\\[-1.8ex]\hline 
\hline \\[-1.8ex] 
 Domain& Variable & In-degree & Out-degree & Mb size \\ 
\hline \\[-1.8ex] 
& AREA & $0$ & $21$ & $28$ \\ 
& YEAR & $0$ & $0$ & $0$ \\ 
\midrule
Health& Life Exp Birth & $0$ & $0$ & $0$ \\ 
 & Life Exp 
 Health & $0$ & $0$ & $0$ \\ 
 & Life Exp 
 65 years & $0$ & $0$ & $0$ \\ 
 \midrule
Education & Nursery School & $0$ & $0$ & $0$ \\ 
 & High School 
  Diploma & $2$ & $3$ & $5$ \\ 
 & University 
 Degree & $1$ & $0$ & $1$ \\ 
 & School 
 Early Leavers  & $1$ & $0$ & $1$ \\ 
 & Lifelong 
  Learning & $1$ & $0$ & $1$ \\ 
   \midrule

 Work \& Life Balance & Employment 
 Rate & $0$ & $4$ & $5$ \\ 
  \scriptsize{EMPLOYMENT} &&\\ \midrule

 Work \& Life Balance & Temporary 
 Jobs & $1$ & $0$ & $1$ \\ 
 \scriptsize{WORK QUALITY} & Low Wage & $1$ & $0$ & $1$ \\ 
 & Not Regular 
 Occup & $0$ & $1$ & $2$ \\ 
& Satisfied 
 Employees & $1$ & $0$ & $1$ \\ 
 & Involuntary 
 Part-Time & $0$ & $1$ & $2$ \\ 
  \midrule
Economic Well-being & Income 
 Per Capita & $2$ & $1$ & $3$ \\ 
\scriptsize{INCOME INEQUALITY} & Income 
 Inequality & $0$ & $0$ & $0$ \\ 
&&\\ \midrule
Economic Well-being & Material 
 Deprivation & $0$ & $0$ & $0$ \\ 
 \scriptsize{MINIMUM ECONOMIC} & Low Quality 
 House & $1$ & $0$ & $1$ \\ 
 \scriptsize{CONDITIONS} & Economic 
 Difficulty & $0$ & $1$ & $1$ \\ 
 & Low Work 
 Intenisty & $1$ & $0$ & $1$ \\ 
  \midrule
Relationships & Family 
 Satisfaction & $0$ & $1$ & $1$ \\ 
 & Friendship 
 Satisfaction & $1$ & $0$ & $1$ \\ 
 & People to Rely 
 Upon & $2$ & $0$ & $2$ \\ 
 & Social 
 Participation & $0$ & $1$ & $2$ \\ 
 & Civic and Political 
 Participation & $0$ & $0$ & $0$ \\ 
 & Volunteering & $1$ & $2$ & $4$ \\ 
 & Finincing 
 Associations & $2$ & $1$ & $3$ \\ 
  & Generalized 
 Trust & $0$ & $0$ & $0$ \\

 \hline \\[-1.8ex] 
\end{tabular} 
\end{table}

\begin{table}[!htbp] \centering 
  \caption{Network estimated by adopting prior Strategy 1 with NUTS 1 level areas (NW-NE-C-S-I): in-degree, out-degree and Markov blanket size of the nodes (Outcome domains).} 
  \label{degree.2} 
  \scriptsize
\begin{tabular}{@{\extracolsep{5pt}} lllll} 
\\[-1.8ex]\hline 
\hline \\[-1.8ex] 
 Domain& Variable & In-degree & Out-degree & Mb size \\ 
 \midrule
Security & Killings & $0$ & $0$ & $0$ \\ 
 \scriptsize{MURDERS} &&\\
\midrule
Security & Home 
 Burglaries & $1$ & $0$ & $1$ \\ 
 \scriptsize{PREDATORY CRIMES} & Pickpocketing & $1$ & $1$ & $4$ \\ 
 & Robberies & $2$ & $0$ & $4$ \\
 \midrule
Subjective Well-being & Life 
 Satisfaction & $0$ & $1$ & $2$ \\ 
 \midrule
Landscape & Cultural  
 Spending & $1$ & $0$ & $1$ \\ 
 & Illegal 
 Buildings & $2$ & $1$ & $3$ \\ 
 & Agritourism 
 Diffusion & $2$ & $0$ & $3$ \\ 
 & Landscape 
 Dissatisfaction & $1$ & $1$ & $2$ \\ 
 \midrule
Environment  & Water Supply 
 Dispersion & $1$ & $1$ & $2$ \\ 
 & Waste 
 to Landfill & $2$ & $0$ & $2$ \\ 
 & Air Quality & $1$ & $0$ & $1$ \\ 
 & Urban Green & $1$ & $2$ & $3$ \\ 
 & Environment 
 Satisfaction & $1$ & $2$ & $4$ \\ 
 & Protected 
 Areas & $2$ & $0$ & $4$ \\ 
 & Renewable 
 Energy & $2$ & $1$ & $4$ \\ 
 & Separate 
 Urban Waste & $1$ & $2$ & $4$ \\ 
 \hline \\[-1.8ex] 
\end{tabular} 
\end{table}

\begin{table}[!htbp] \centering 
  \caption{Network estimated by adopting prior Strategy 1 with NUTS 1 level areas (NW-NE-C-S-I): in-degree, out-degree and Markov blanket size of the nodes (Context domains).}  
  \label{degree2} 
  \scriptsize
\begin{tabular}{@{\extracolsep{5pt}} lllll} 
\scriptsize
\\[-1.8ex]\hline 
\hline \\[-1.8ex] 
 Domain& Variable & In-degree & Out-degree & Mb size \\ 
\hline \\[-1.8ex]

Politics  & Parliament 
 Trust & $0$ & $1$ & $2$ \\ 
 & Judicial System 
 Trust & $1$ & $0$ & $1$ \\ 
 & Parties Trust & $2$ & $1$ & $3$ \\ 
 & Other Institutions 
 Trust & $0$ & $0$ & $0$ \\ 
 & Women Political 
 Representation & $0$ & $1$ & $2$ \\ 
 & Duration Civil 
 Proceedings & $2$ & $0$ & $2$ \\ 
 & Prison 
 Overcrowding & $0$ & $0$ & $0$ \\ 
 \midrule
Research & Research 
 Intensity & $2$ & $0$ & $4$ \\ 
 & Knowledge 
  Workers & $0$ & $1$ & $2$ \\ 
 & Creativity 
 Workers & $1$ & $0$ & $1$ \\ 
 \midrule
Services & Beds in 
 Hospitals & $2$ & $0$ & $2$ \\ 
 & Childcare 
 Services & $2$ & $0$ & $2$ \\ 
 & Services Access 
  Difficulty & $0$ & $0$ & $0$ \\ 
 & Water 
 Irregularities & $2$ & $0$ & $2$ \\ 
 & Places-km & $1$ & $2$ & $3$ \\ 
 & Mobility 
 Satisfaction & $1$ & $0$ & $1$ \\ 
\hline \\[-1.8ex] 
& Average & 0.85 &  0.89 & 2.16\\
& St. Dev. & 0.78 &  2.7 & 3.64\\
\bottomrule
\end{tabular} 
\end{table}

\begin{table}
  \centering
  \caption{Network estimated by adopting prior Strategy 1 with NUTS 1 level areas (NW-NE-C-S-I): direct connections among domains.}
 \label{strategy2}
\scriptsize
\begin{threeparttable}[b]
 
  \begin{tabular}{lll}
\\[-1.8ex]\hline 
\hline \\[-1.8ex] 
 Domain &    \multicolumn{2}{l}{Directly Connected Domains}\\ 
 \cmidrule{2-3}\\
 & Number& Domains\tnote{1}  \\
 \midrule
 Health & 0 & \\
 \midrule
 Education & 2 & Politics, Research\\
 \midrule
 Work \& Life Balance & 1 & Economic Well-being \scriptsize{(INCOME AND INEQUALITY)}, \\

\midrule
   Work \& Life Balance & 4 & Economic Well-being  \scriptsize{(INCOME AND INEQUALITY)}, \\ 
   && Relationships, Services, Research \\
  \midrule
 Economic Well-being &2 & Work \& Life Balance  \scriptsize{(EMPLOYMENT)}, \\
 \scriptsize{(INCOME AND INEQUALITY)}  &&Work \& Life Balance   \scriptsize{(QUALITY)}  \\
 \midrule
 Economic Well-being & 0 & \\
  \scriptsize{(MIN. ECONOMIC CONDITIONS)} &\\
   \midrule
   Relationships & 2 & Environment, Work \& Life Balance  \scriptsize{(QUALITY)}\\
      \midrule
      Security & 0 &\\
 \scriptsize{(MURDERS)} &\\
    \midrule
    Security & 1 & Environment \\
   \scriptsize{(PREDATORY CRIMES)} &\\
   \midrule
   Subjective Well-being & 1 & Politics \\
  \midrule
  Landscape & 2 & 
   Environment (3), Services  \\
   \midrule
   Environment & 4 &  Landscape (3), Relationships, Services, \\
   &&Security \scriptsize{(PREDATORY CRIMES)} \\
   \midrule
   Politics & 3 & Education, Subjective Well-being,  Research\\
   \midrule
   Research & 3 & Education, Work \& Life Balance  \scriptsize{(QUALITY)}, Politics \\
   \midrule
   Services & 3 & Work \& Life Balance  \scriptsize{(QUALITY)}, Environment,\\ && Landscape\\
   \bottomrule
\end{tabular}
\begin{tablenotes}
\item [1] Notice that if two domains have more than one direct connection, than the exact number of direct connections is put in parenthesis
\end{tablenotes}
\end{threeparttable}
\end{table}

\clearpage
\subsection*{References}

\renewcommand{\bibsection}{}
\bibliography{references}
 
\bigskip 

\bigskip

\small {\bf Authors' address:}

\bigskip 

\noindent

\noindent University of Rome La Sapienza, 
\hfill {\tt onori.federica@gmail.com}\\
Faculty of 
Information Engineering,    Informatics and Statistics, \\
Department of Statistical Sciences, \hfill {\tt giovanna.jonalasinio@uniroma1.it}\\
Italy

\end{document}